\documentclass[chicago]{emulateapj}

\newcommand{\SQlong}{2003$\,$SQ$_{317}$} 
\newcommand{\SQ}{SQ$_{317}$} 
\newcommand{\QG}{QG$_{298}$} 

\shorttitle{The Unusual KBO \SQlong}
\shortauthors{Lacerda et al.}

\begin{document}

\title{The Unusual Kuiper Belt Object \SQlong}

\author{Pedro Lacerda\altaffilmark{1}, Andrew McNeill}
\affil{Astrophysics Research Centre, Queen's University
Belfast, Belfast BT7 1NN}

\and

\author{Nuno Peixinho\altaffilmark{2}} 
\affil{Center for Geophysics of the University of Coimbra\\
Geophysical and Astronomical Observatory of the University of
Coimbra\\ Almas de Freire, 3040-004 Coimbra, Portugal}

\altaffiltext{1}{lacerda.pedro@gmail.com}
\altaffiltext{2}{Unidad de Astronom\'{\i}a, Universidad de Antofagasta, Avenida
Angamos 601, Antofagasta, Chile}

\slugcomment{Submitted to MNRAS}

\begin{abstract}

  We report photometric observations of Kuiper belt object \SQlong\
  obtained between 2011 August 21 and 2011 November 1 at the 3.58 m
  New Technology Telescope, La Silla. We obtained a rotational
  lightcurve for \SQlong\ with a large peak-to-peak photometric range,
  $\Delta m=0.85\pm0.05$ mag, and a periodicity, $P=7.210\pm0.001$ hr.
  We also measure a nearly neutral broadband colour $B-R=1.05\pm0.18$
  mag and a phase function with slope $\beta=0.95\pm0.41$ mag/$\degr$.
  The large lightcurve range implies an extremely elongated shape for
  \SQlong, possibly as a single elongated object but most simply
  explained as a compact binary. If modelled as a compact binary near
  hydrostatic equilibrium, the bulk density of \SQlong\ is near 2670
  kg m$^{-3}$. If \SQlong\ is instead a single, elongated object, then
  its equilibrium density is about 860 kg m$^{-3}$. These density
  estimates become uncertain at the 30\% level if we relax the
  hydrostatic assumption and account for solid, ``rubble pile''-type
  configurations. \SQlong\ has been associated with the Haumea family
  based on its orbital parameters and near-infrared colour; we discuss
  our findings in this context.  If confirmed as a close binary,
  \SQlong\ will be the second object of its kind identified in the
  Kuiper belt.    

\end{abstract}


\section{Introduction}

Haumea is a large, triaxial KBO (semi-axes
$2000\times1600\times1000$ km), with a very fast rotation (period
$P\approx3.9$ hr), a rock-rich interior (bulk density
$\rho\approx2500$ kg m$^{-3}$) and a surface covered in high-albedo
($p\approx0.8$), nearly pure water ice, which shows signs of
variegation \citep{2006ApJ...639.1238Rabinowitz,
  2007AJ....133.1393Lacerda, 2010A&A...518L.147Lellouch,
  2007ApJ...655.1172T, 2008AJ....135.1749Lacerda,
2009AJ....137.3404Lacerda}. Haumea has two, nearly coplanar satellites
with similarly icy surfaces \citep{2005ApJ...632L..45Brown,
2006ApJ...640L..87Brown, 2011A&A...528A.105Dumas}. 

At least 10 other KBOs have been associated with Haumea on the basis
that they have similar orbital elements and water-ice-rich surfaces
\citep{2007Natur.446..294Brown, 2007AJ....134.2160Ragozzine,
  2008ApJ...684L.107Schaller, 2010A&A...511A..72Snodgrass,
2012A&A...544A.137Carry}. The origin of this so-called Haumea family
is unclear. Proposed, ad-hoc scenarios include a giant impact onto
the proto-Haumea \citep{2007Natur.446..294Brown}, a gentler
graze-and-merge collision \citep{2010ApJ...714.1789Leinhardt} and
a sequence of two collisions in which the first creates a moon
which is the target of the second collision
\citep{2009ApJ...700.1242Schlichting}. The first scenario is ruled
out by the low velocity dispersion of the family members, while
the last two possibilities are arguably improbable. Furthermore,
the mass in the currently known family members and their velocity
dispersion is not ideally matched by any of the proposed scenarios
\citep{2012Icar..221..106Volk}.

\citet{2010A&A...511A..72Snodgrass} noticed that one member of the
Haumea family, \SQlong, displayed large photometric variation,
$\sim$1 mag peak-to-peak, in just 14 measurements. They estimated a
periodicity of about 3.7 hr (or twice that) for this object. Such
large variability in 200 km-scale objects often indicates extreme
shapes from which useful information can be extracted
\citep{1978Icar...36..353H, 1980Icar...44..807Weidenschilling,
2004AJ....127.3023S, 2007AJ....133.1393Lacerda}.

Here we report time-resolved, follow-up observations of \SQlong\
(hereafter \SQ) obtained to clarify the nature of this object and
the cause for the extreme variability and to improve our
understanding of the Haumea family. We find that \SQ\ indeed has an
extreme shape, most simply explained by a compact binary, although
more data are needed to rule out a single, elongated shape.

\section[]{Observations} \label{sec:Observations}

\begin{deluxetable*}{lcccclll}
\tabletypesize{\small}
\tablewidth{0pt}
\tablecaption{Journal of Observations. \label{Table.Journal}}
\tablehead{
  \colhead{Date UT} & \colhead{$R$} & \colhead{$\Delta$} &
  \colhead{$\alpha$} & \colhead{Seeing} & \colhead{Filter} &
  \colhead{Exposure} & \colhead{Conditions} \\
  \colhead{} & \colhead{[AU]} & \colhead{[AU]} & \colhead{[\degr]} &
  \colhead{[\arcsec]} & \colhead{} & \colhead{[sec]} & \colhead{}} 
  \startdata                                    
    2011 Aug 21 & 39.2440 & 38.4556  & 0.941    &1.0       & R & 600
    & photometric \\
    2011 Aug 22 & 39.2440 & 38.4452  & 0.921    &0.9       & R & 420
    & photometric \\
    2011 Aug 23 & 39.2440 & 38.4351  & 0.901    &1.7       & R,B &
    420, 600 & photometric \\
    2011 Oct 30 & 39.2433 & 38.3911  & 0.748    &0.8       & R & 300
    & photometric \\
    2011 Oct 31 & 39.2432 & 38.4003  & 0.769    &0.7       & R
    & 300      & photometric \\ 
    2011 Nov 01 & 39.2432 & 38.4097  & 0.790    &0.8       & R
    & 300      & thin cirrus 
  \enddata
  \tablecomments{Columns are (1) UT date of observations, (2)
  heliocentric distance to KBO, (3) geocentric distance to KBO, (4)
  solar phase angle, (5) atmospheric seeing, (6) filters used, (7)
  exposure times used, and (8) atmospheric conditions.}
\end{deluxetable*}

\begin{deluxetable*}{lcccc}
\tabletypesize{\small}
\tablewidth{0pt}
\tablecaption{Photometry. \label{Table.Photometry}}
\tablehead{
  \colhead{Date UT} & \colhead{$m_R$} & \colhead{$m_B$} &
  \colhead{$B-R$} & \colhead{$m_R(1,1,\alpha)$} \\ 
  \colhead{} & \colhead{[mag]} & \colhead{[mag]} & \colhead{[mag]} &
  \colhead{[mag]}}
  \startdata
  2011 Aug 21 & $22.44\pm0.10$ & \nodata       & \nodata      & $6.55\pm0.10$ \\
  2011 Aug 22 & $22.25\pm0.03$ & \nodata       & \nodata      & $6.36\pm0.03$ \\
  2011 Aug 23 & $22.29\pm0.04$ & $23.34\pm0.18$ & $1.05\pm0.18$ & $6.40\pm0.04$ \\
  2011 Oct 31 & $22.13\pm0.04$ & \nodata       & \nodata      & $6.24\pm0.04$ \\ 
  2011 Nov 01 & $22.23\pm0.15$ & \nodata       & \nodata      & $6.34\pm0.15$ \\ 
  \enddata
  \tablecomments{Columns are (1) UT date of observations, (2) apparent
  $R$ magnitude, (3) apparent $B$ magnitude, (4) $B-R$ colour, and (5)
  absolute $R$ magnitude, uncorrected for illumination phase
  darkening. All magnitudes are at maximum lightcurve flux.}
\end{deluxetable*}

We observed KBO \SQ\ using the 3.58m ESO New Technology Telescope
(NTT) located at the La Silla Observatory, in Chile. The NTT was
configured with the EFOSC2 instrument
\citep{1984Msngr..38....9Buzzoni, 2008Msngr.132...18Snodgrass} mounted
at the f/11 Nasmyth focus and equipped with a LORAL $2048\times2048$
CCD. We used the $2\times2$ binning mode bringing the effective pixel
scale to 0.24\arcsec/pixel.  Our observations were taken through
Bessel $B$ and $R$ filters (ESO \#639 and \#642).

Each night, we collected bias calibration frames and dithered, evening
and morning twilight flats through both filters. Bias and flatfield
frames were grouped by observing night and then median-combined into
nightly bias, and $B$ and $R$ flatfields. The science images were also
grouped by night and by filter and reduced (bias subtraction and
division by flat field) using the IRAF {\tt ccdproc} routine.  The $R$
band images suffered from slight fringing which was removed using an
IRAF package optimised for EFOSC2 \citep{2013Msngr.152...14Snodgrass}. 

On photometric nights, we used observations of standard stars (MARK
A1-3, 92~410, 94~401, PG2331+055B) from \citet{1992AJ....104..340Lan}
to achieve absolute calibration of field stars near \SQ. We employed
differential photometry relative to the calibrated field stars to
measure the magnitude of \SQ\ as a function of time. Uncertainty in
the differential photometry of \SQ\ (typically $\pm0.05$ mag) was
estimated from the dispersion in the measurements relative to
different stars.

Table 1 presents a journal of the observations and Table 2 lists the
calibrated, apparent magnitudes at peak brightness, measured for \SQ\
on photometric nights.

\section{Results} \label{sec:Discussion}

\begin{deluxetable*}{@{}rcl@{}}
\tablewidth{0pt}
\tablecaption{Properties of \SQlong \label{Table.Properties}}
\tablehead{\colhead{Property} & \colhead{Symbol} & \colhead{Value}}
\startdata
Orbital semimajor axis & $a$ & 42.753 AU \\
Orbital eccentricity & $e$ & 0.082 \\
Orbital inclination & $i$ & 28.6\degr \\
Equiv.\ diameter ($p=0.50$) & $D$ & 150 km \\
Equiv.\ diameter ($p=0.05$) & $D$ & 470 km \\
Absolute magnitude & $m_R(1,1,0)$  & $5.52\pm0.36$ mag \\
Phase function slope & $\beta$       & $0.95\pm0.41$ mag/$\degr$\\
Lightcurve period & $P$           & $7.210\pm0.001$ hr \\
Lightcurve variation & $\Delta m$    & $0.85\pm0.05$ mag 
  \enddata
\tablecomments{
Equivalent diameter is calculated from the measured absolute
magnitude for two possible values of the geometric albedo using
$D=(1329\,\mathrm{km})\,p^{-0.5}\,10^{-0.2\,m_R(1,1,0)}$}.
\end{deluxetable*}

\begin{figure*}
  \includegraphics[width=160mm]{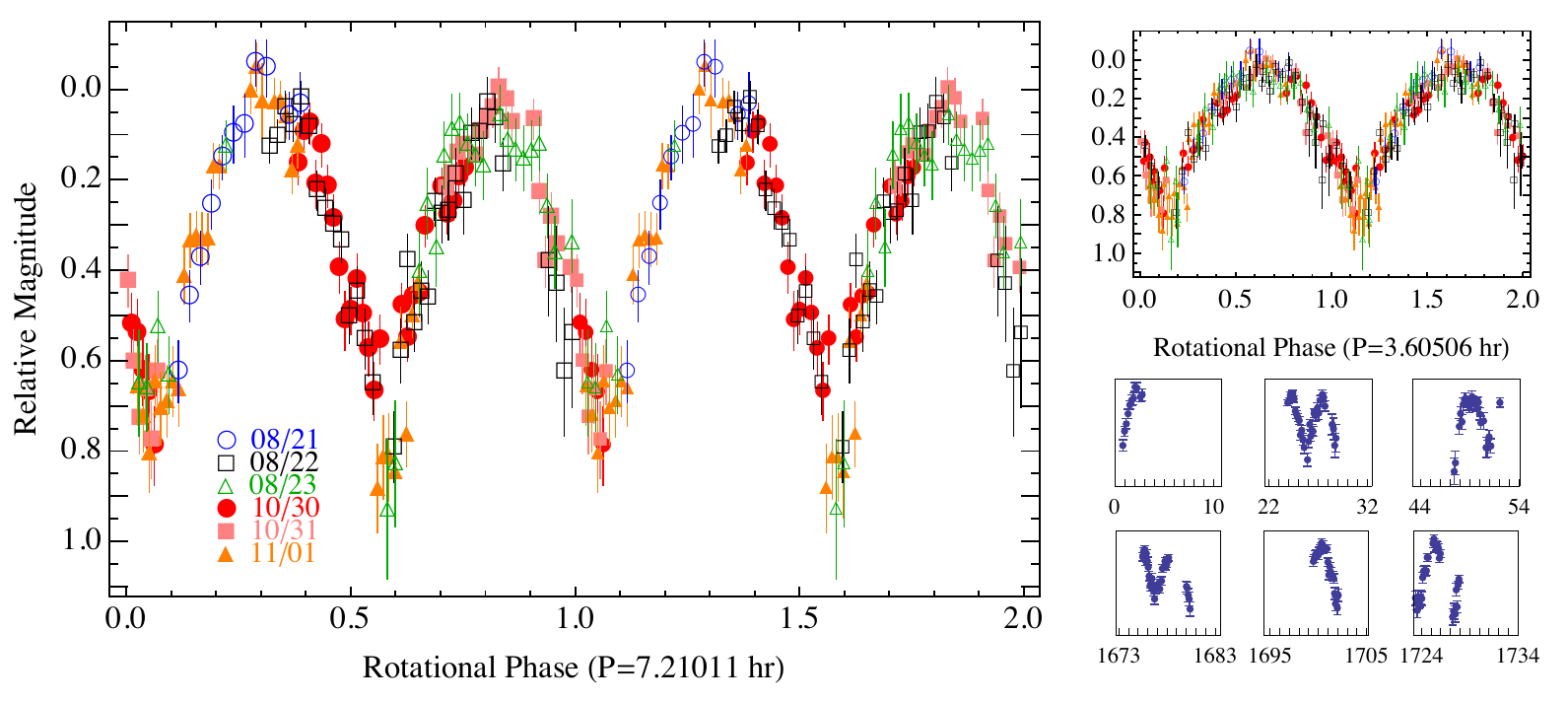}
  \centering

  \caption{Lightcurve of \SQ. Large panel shows data phased with the
    best-fit spin period $P=7.21011$ hr. Two full rotations are shown.
    Smaller panels on the right show: (top) lightcurve phased with the
    possible but less likely period of $\sim$3.6 hr, and (bottom)
    measurements on each individual night with the $x$-axis (time)
    labeled in hours since MJD 55794.0 and the $y$-axis (unlabeled)
    equal to that of the figures above and to the left. The times are
    light-travel-time subtracted and so indicate when the light left
  \SQ.}

  \label{fig:Lightcurves}
\end{figure*}

\begin{figure*}
  \includegraphics[width=160mm]{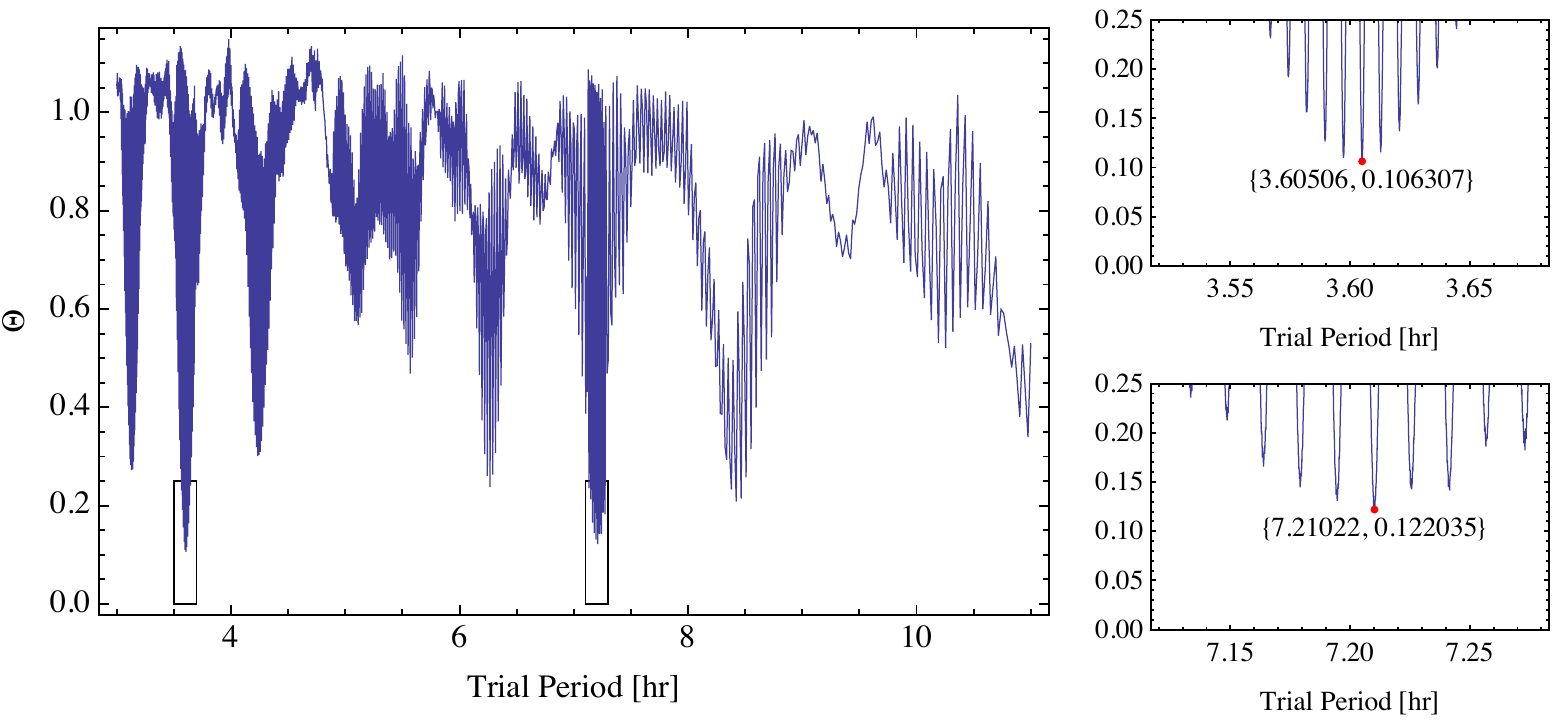}
  \centering

  \caption{Phase dispersion minimization periodogram for the \SQ\
    data. Minima of the quantity $\Theta$ mark the most likely
    rotational periods. The two deepest minima are marked with boxes
    and enlarged on the right; the best-fit periods are shown in the
    insets together with the respective $\Theta$ value. Aliasing
    resulting from the way the data were sampled is visible in the 
    insets.}

  \label{fig:PDMPlot}
\end{figure*}

\begin{figure}
  \includegraphics[width=83mm]{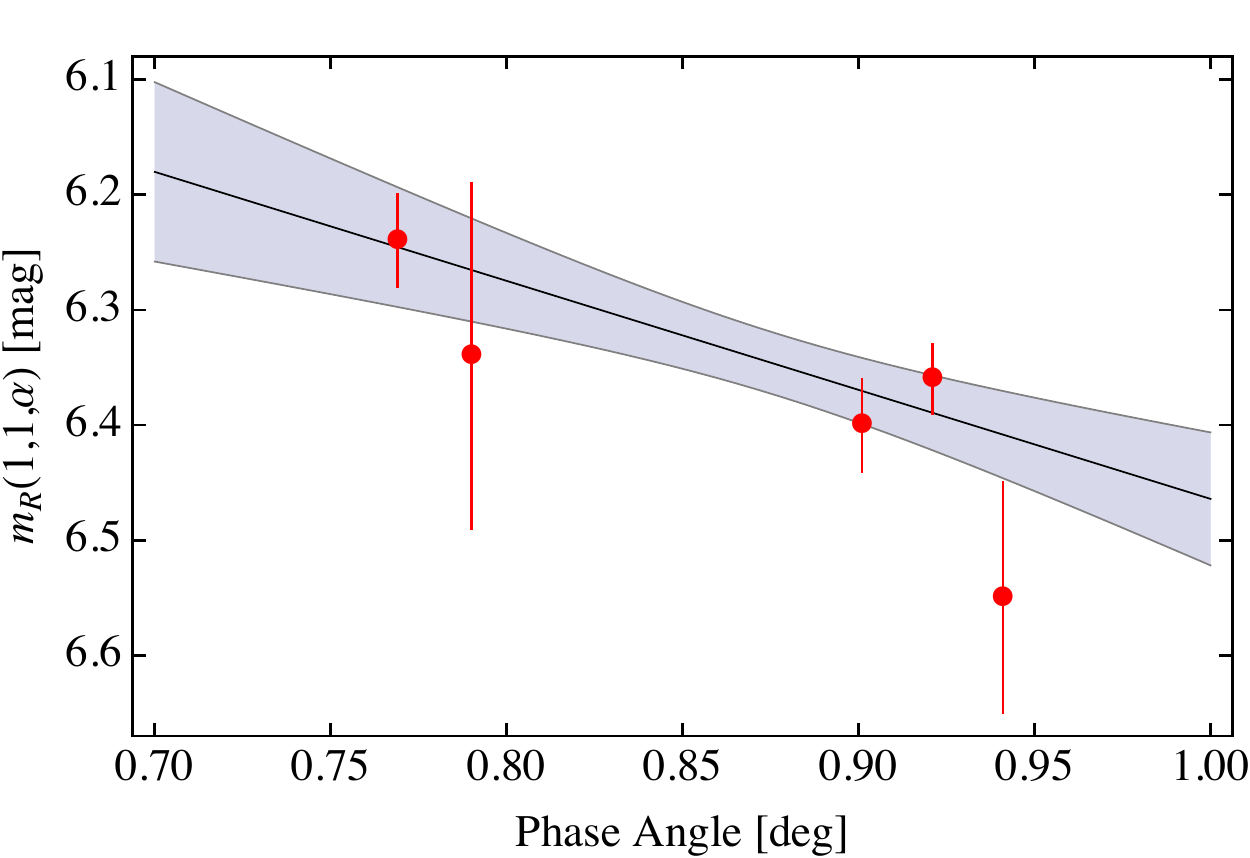}
  \centering

  \caption{Phase curve of \SQ\ as inferred from five measurements
    (filled circles with error bars). A least-squares fit to the data
    is indicated as a black line within a shaded, 1-$\sigma$
    confidence region. The best fit has a slope $\beta=0.96\pm0.41$
  mag/$\degr$ and intercept $m_R(1,1,0)=5.52\pm0.36$ mag.}

  \label{fig:PhaseCurve}
\end{figure}

\subsection{Rotational Lightcurve} \label{sec:Lightcurve}

Our photometry of \SQ\ resulted in $N=154$ measurements over 6 nights
spanning a total interval of $T=1728$ hours (Figure
\ref{fig:Lightcurves}), The brightness of \SQ\ varies visibly, by
$\Delta m=0.85\pm0.05$ mag peak-to-peak, taking only 1.8 hours to go
from minimum to maximum brightness. The full extent of this variation
is seen on multiple nights. 

To search for periodicity in the data we employed two methods: the
Phase Dispersion Minimization \citep[PDM;][]{1978ApJ...224..953S} and
the String-Length Minimization
\citep[SLM;][]{1983MNRAS.203..917Dworetsky}. PDM minimises the ratio,
$\Theta$, between the scatter of the data phased with a trial period
and that of the unphased data. The best-fit period will result in a
lightcurve with the least scatter, hence minimising $\Theta$. SLM
minimises the length of a segmented line connecting the data points
phased with a trial period. Similarly to PDM, the best-fit period will
result in a lightcurve with the smallest scatter around the real,
periodic lightcurve and hence the shortest string length. Before
running the period-search algorithms we corrected the observing times
by subtracting the light-travel time from \SQ\ to Earth for each
measurement.

Figure \ref{fig:PDMPlot} shows the PDM periodogram for lightcurve
periods ranging from 3 to 11 hours. Periods outside this range
resulted in larger values of $\Theta$. Two strong PDM minima are
apparent, one at $P_{1/2}\approx3.6$~hr implying a single-peaked
lightcurve with one maximum and one minimum per full rotation, and
another at $P=2\times P_{1/2}\approx7.2$~hr which folds the data onto
a double-peaked lightcurve (Figure \ref{fig:Lightcurves}).

We favour the double-peaked solution, $P\approx7.2$~hr, for three
reasons. Firstly, a single-peaked lightcurve with a variation
$\sim0.85$ mag would have to be caused by a peculiar, large contrast,
surface albedo pattern. The symmetry and regularity of the lightcurve
suggest that the brightness variation is modulated instead by the
elongated shape of \SQ\ as it rotates; lightcurves produced by shape
are double peaked. Secondly, the double-peaked solution produces a
lightcurve with slightly asymmetric minima, seen on more than one
night. The single-peaked lightcurve minimum exhibits more scatter
suggesting that it is a superposition of two different minima.
Finally, the single-peaked period, $P_{1/2}\approx3.6$~hr, would imply
very fast rotation, at which \SQ\ would likely experience significant
centripetal deformation for a plausible range of bulk densities and
inner structures. The resulting elongated shape would produce a
double-peaked lightcurve invalidating the premise that the lightcurve
is single-peaked.

High resolution analysis near the $7.2$~hr lightcurve indicates a PDM
minimum at $P_\mathrm{PDM}=7.21022$~hr, while using the SLM method, we
obtained a best-fit period $P_\mathrm{SLM}=7.20999$~hr. We take as
best-fit solution the mean of the two, $P=7.21011$~hr. To estimate the
uncertainty in our period solution we employ \citep{1986ApJ...302..757Horne}

\[
  \delta f=\frac{3\pi\,\sigma_N}{2\sqrt{N}\,T\,\Delta m}
\]

\noindent where $\delta f$ is the uncertainty in the lightcurve
frequency, $\sigma_N$ is the standard deviation of the lightcurve
best-fit residuals (calculated using the model shown in Figure
\ref{fig:BinaryLightcurveFit}), $N=154$ is the number of data points,
$T=1728$~hr is the total time spanned by the observations, and $\Delta
m=0.85$~mag is the lightcurve variation. The frequency uncertainty is
$\delta f=0.00002$ hr$^{-1}$ which corresponds to an uncertainty in
the spin period $\delta P=0.00103$~hr. We therefore adopt as best
period $P=7.210\pm0.001$~hr.

\subsection{Phase Curve} \label{seq:PhaseCurve}

Owing to their large heliocentric distances, Kuiper belt objects are
only observable from Earth at small phase angles, $\alpha<1.5\degr$.
Our observations span an approximate range $0.75<\alpha\,
[\degr]<0.95$ and we see a trend of fainter apparent magnitude
with increasing phase angle.  We fitted this observed phase darkening
with a weighted linear model of the form

\[
  m_R(1,1,\alpha)=m_R(1,1,0)+\beta\,\alpha
\]

\noindent where $m_R(1,1,0)$ is the absolute magnitude at zero phase
angle and $\beta$ is the linear phase curve coefficient. The fit is
plotted in Figure \ref{fig:PhaseCurve} where we show apparent
magnitudes at lightcurve maxima. The best-fit zeropoint and slope are
$m_R(1,1,0)=5.52\pm0.36$ mag and $\beta=0.95\pm0.41$ mag/$\degr$. The
phase function slope is steeper (although only by 2$\sigma$) than what
is typically seen in other KBOs \citep[$\beta_\mathrm{KBO}\sim0.16$
mag/$\degr$;][]{2002AJ....124.1757Sheppard, 2007AJ....133...26R} and
does not follow the trend for shallower phase functions observed in
other objects associated with Haumea
\citep{2008AJ....136.1502Rabinowitz}. We note that the phase function
found above is consistent with the measurement $m_R=22.05\pm0.02$ mag
on 2008/08/30 (at phase angle $\alpha=0.62\degr$, heliocentric
distance $r=39.261$~AU and geocentric distance $\Delta=38.342$~AU) by
\citet{2010A&A...511A..72Snodgrass}. However, because of the narrow
range of phase angles sampled, the uncertainty in the phase function
slope is large so we are reluctant to draw strong implications from
this result.

\subsection{Shape Model} \label{sec:Shape}

\begin{deluxetable*}{cccccccc}
\tabletypesize{\small}
\tablewidth{0pt}
\tablecaption{Model Fit Parameters \label{table:FitParameters}}
\tablehead{\colhead{Model Type} & \colhead{$q$} & \colhead{$d$} &
\colhead{$B/A$} & \colhead{$C/A$} & \colhead{$b/a$} & \colhead{$c/a$}
& \colhead{$\rho$ [kg m$^{-3}$]}}
      \startdata
  Jacobi Ellipsoid & \ldots                 & \ldots
                   & $0.55^{+0.05}_{-0.04}$ & $0.41^{+0.02}_{-0.02}$
                   & \ldots                 & \ldots
                   & $861^{+33}_{-32}$ \\
      Roche Binary & $0.32^{+0.08}_{-0.07}$ & $1.14^{+0.14}_{-0.04}$
                   & $0.91^{+0.01}_{-0.03}$ & $0.82^{+0.01}_{-0.02}$
                   & $0.51^{+0.15}_{-0.07}$ & $0.48^{+0.13}_{-0.07}$
                   & $2671^{+88}_{-102}$ \\
      \enddata
    \tablecomments{
    Columns are (1) Model used to fit lightcurve, (2) component mass
    ratio, (3) binary separation in units of $A+a$, (4) and (5)
    primary semimajor axes, (6) and (7) secondary semimajor axes, and
    (8) model bulk density.}
\end{deluxetable*}

The large photometric variability of \SQ\ suggests that the object has
a highly elongated, possibly binary shape. Indeed, assuming that \SQ\
is close to hydrostatic equilibrium, its photometric range ($\Delta
m=0.85$ mag) and spin frequency ($\omega=3.33$ day$^{-1}$) place it
near the threshold between the Jacobi ellipsoid and the Roche binary
sequences \citep{1984A&A...140..265L,2004AJ....127.3023S}. To explore
this issue further, we attempt to fit the lightcurve of \SQ\ using
Jacobi ellipsoid and Roche binary hydrostatic equilibrium models. The
choice of models of hydrostatic equilibrium is physically based and
has the benefit of allowing the density of \SQ\ to be estimated. 

We follow the procedure detailed in \cite{2007AJ....133.1393Lacerda}
which considers a grid of models spanning a range of Jacobi ellipsoid
shapes, and Roche binary shapes, mass ratios and separations
calculated using the formalism in \citet{1963ApJ...138.1182C}. Each
model is rendered at multiple rotational phases to extract the
lightcurve. Surface scattering is modelled as a linear combination of
the Lambert and Lommel-Seeliger laws. The former mimics a perfectly
diffuse surface and adequately describes a high-albedo, icy object
displaying significant limb darkening. The latter is meant to simulate
a low albedo, lunar-type surface with negligible limb darkening. These
laws are linearly combined through a parameter, $k$, that varies
between 0 (pure Lommel-Seeliger, lunar-type scattering) and 1 (pure
Lambertian, icy-type scattering). The result is a collection of model
lightcurves that can be compared to the one in Figure
\ref{fig:Lightcurves} to identify the best-fitting model. 

As described in \citet{2007AJ....133.1393Lacerda}, the Jacobi
ellipsoid model lightcurves are fully defined by the model's triaxial
shape (semi-axes $A,B,C$) in terms of the axis ratios $B/A$ and $C/A$,
and by the coefficient $k$. The Roche binary lightcurves are entirely
described by the binary component mass ratio $q$, the primary triaxial
shape defined by the axes ratios $B/A$ and $C/A$, the secondary shape
equally defined by the triaxial axis ratios $b/a$ and $c/a$, the
binary separation, $d\geq1$ (expressed in units of the sum of the
primary and secondary semi-axes $A+a$), and the scattering parameter,
$k$. Roche binaries are assumed to be tidally locked with the
components aligned along their longest axes.

For simplicity and to keep the problem tractable we consider only
models viewed equator-on. By allowing the observing geometry to vary
as a free parameter we would increase the number of models that can
match the lightcurve of \SQ\ and hence the overall degeneracy of the
fitting procedure. Generally, off-equator geometries lead to slightly
larger mass ratio solutions, but this has been shown not to have a
significant effect on the inferred bulk density
\citep{2007AJ....133.1393Lacerda}, arguably the most important derived
property. 

\begin{figure}
  \includegraphics[width=\columnwidth]{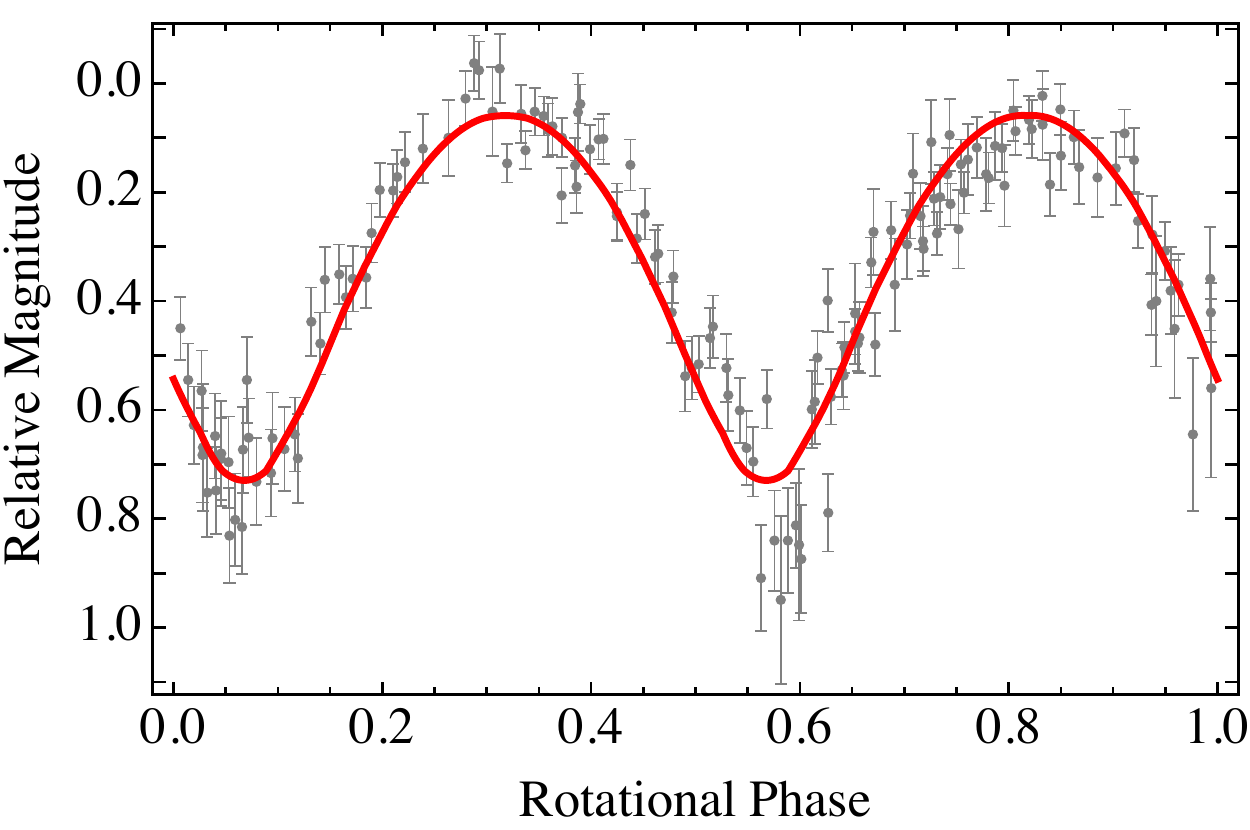}
  \centering

  \caption{Best-fit Jacobi ellipsoid lightcurve (solid line)
  plotted over the lightcurve data for \SQ\ (grey points). The
  corresponding Jacobi ellipsoid is shown in Figure
\ref{fig:EllipsoidModel}.}

  \label{fig:EllipsoidLightcurveFit}
\end{figure}

\begin{figure}
  \includegraphics[width=77mm]{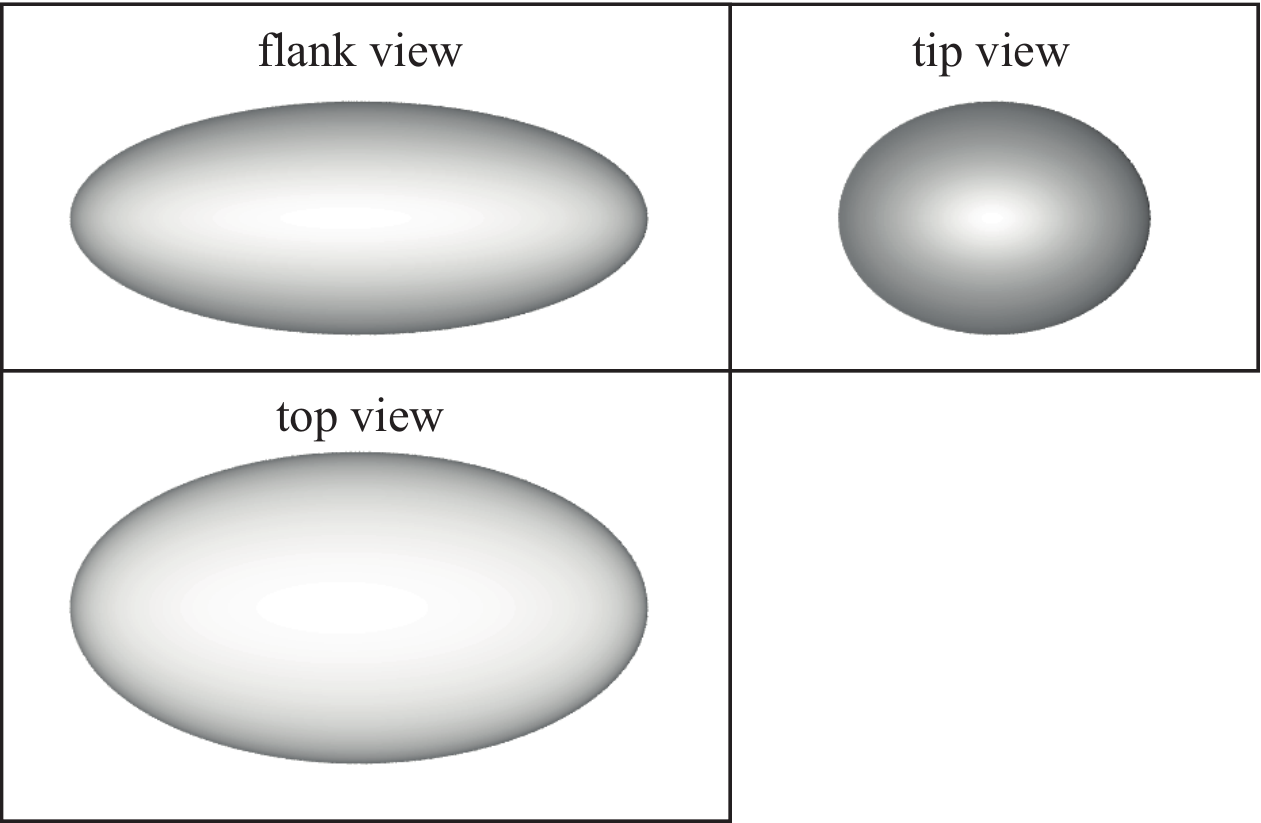}
  \centering

  \caption{Jacobi ellipsoid model that best fits the lightcurve of
  \SQ. }

  \label{fig:EllipsoidModel}
\end{figure}

\begin{figure}
  \includegraphics[width=83mm]{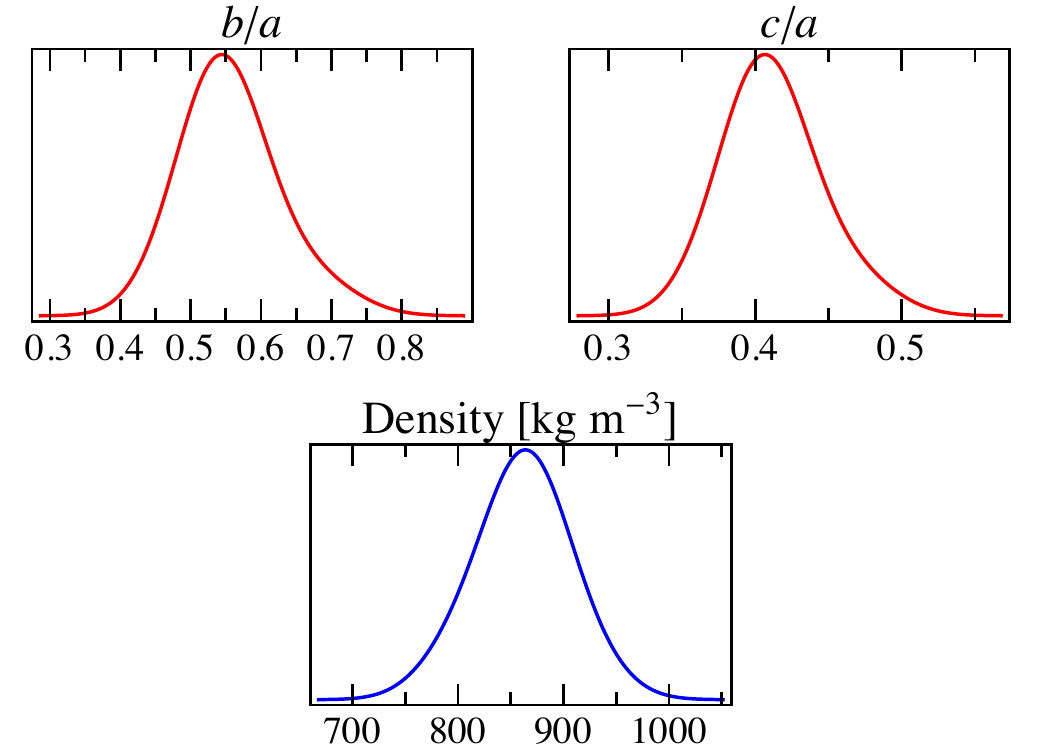}
  \centering

  \caption{Monte Carlo distribution of Jacobi ellipsoid parameters
  ($b/a$ and $c/a$) that fit the lightcurve of \SQ. The Monte Carlo
distribution of bulk density for these models is also shown.}

  \label{fig:EllipsoidParameterDistribution}
\end{figure}

\begin{figure}
  \includegraphics[width=\columnwidth]{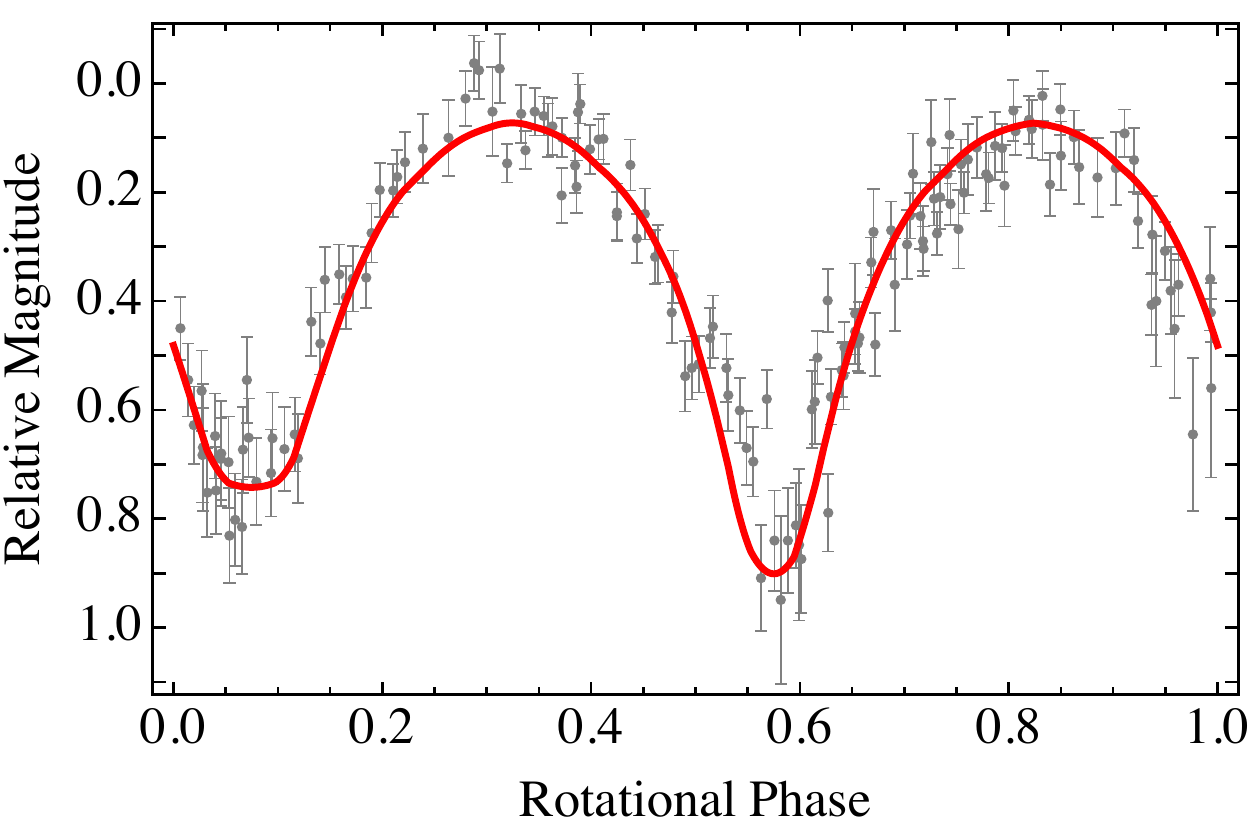}
  \centering

  \caption{Best-fit Roche binary lightcurve (solid line)
  plotted over the lightcurve data for \SQ\ (grey points). The
  corresponding Roche binary is shown in Figure
\ref{fig:BinaryModel}.}

  \label{fig:BinaryLightcurveFit}
\end{figure}

\begin{figure}
  \includegraphics[width=\columnwidth]{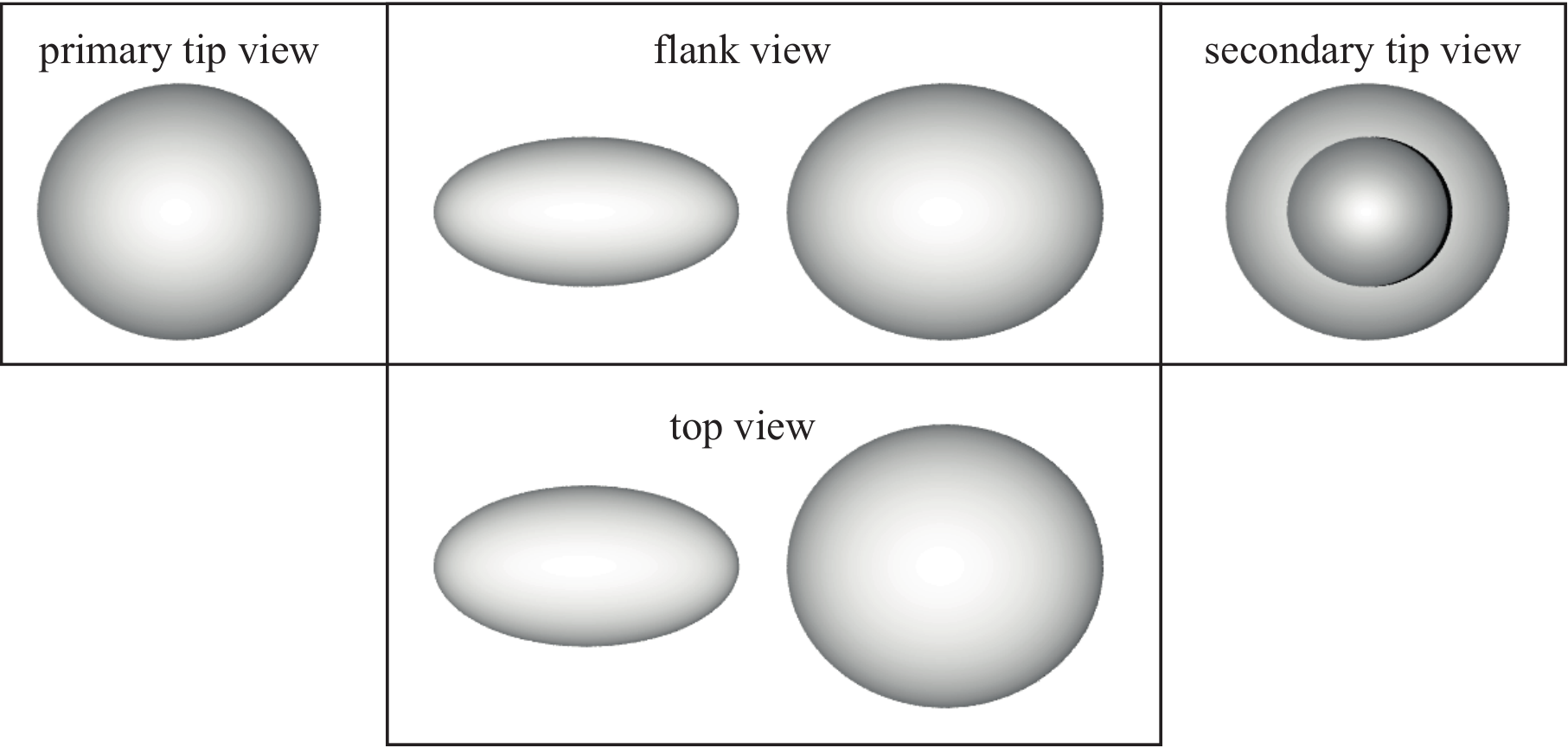}
  \centering

  \caption{Roche binary model that best fits the lightcurve of
  \SQ. }
8800
  \label{fig:BinaryModel}
\end{figure}

\begin{figure}
  \includegraphics[width=77mm]{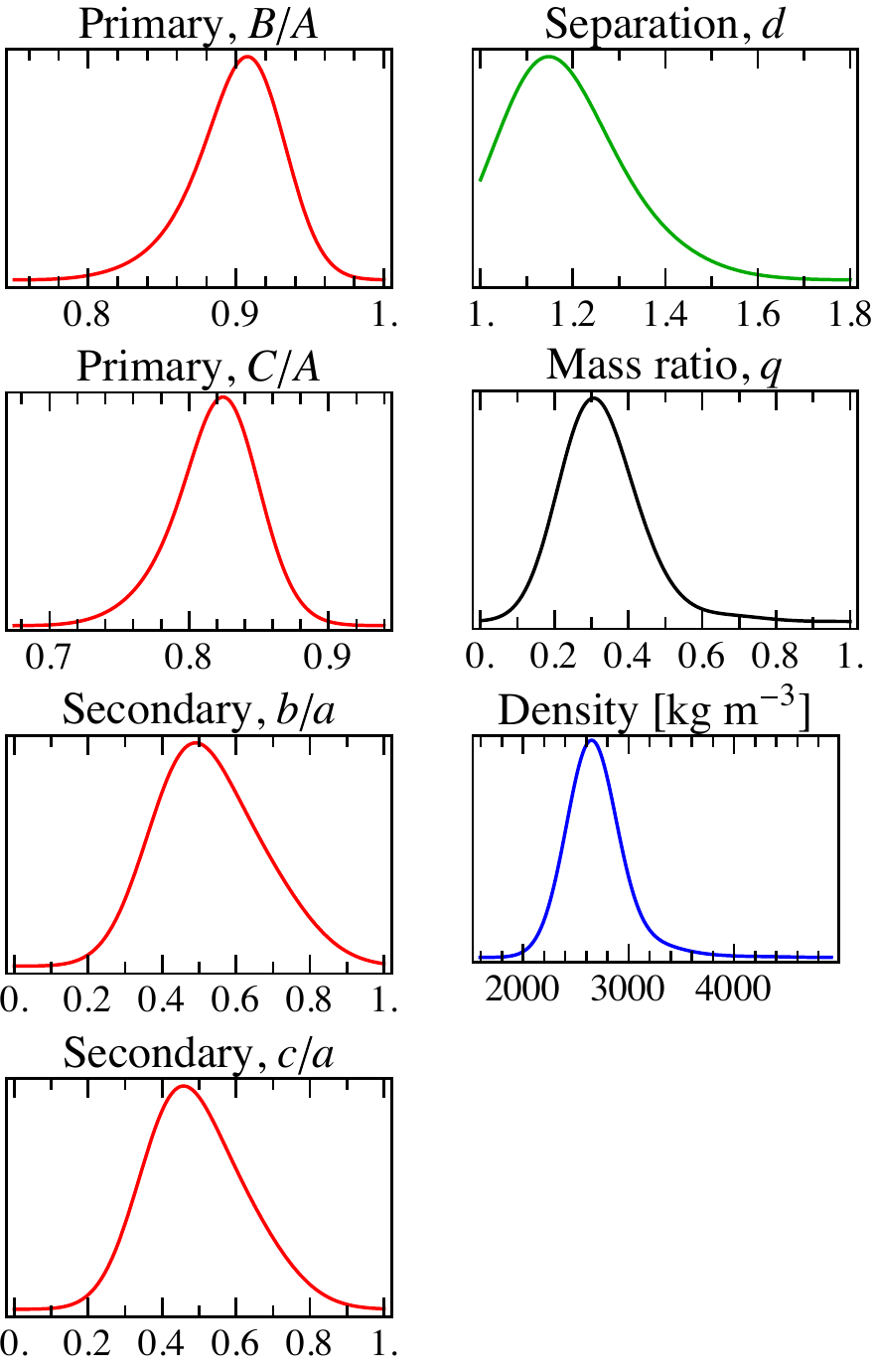}
  \centering

  \caption{Monte Carlo distribution of Roche binary parameters (mass
  ratio $q$, binary separation $d$, $B/A$, $C/A$, $b/a$ and $c/a$)
that fit the lightcurve of \SQ. The Monte Carlo distribution of bulk
density for these models is also shown.}

  \label{fig:BinaryParameterDistribution}
\end{figure}

Each model lightcurve is adjusted (in phase and offset) to the data
using a Levenberg-Marquardt algorithm and the best-fitting one is
selected using a $\chi^2$ criterion. To explore the dependence of this
procedure on the measurement uncertainties we employ a Monte Carlo
approach: we generate $N=371$ bootstrapped instances of the lightcurve
of \SQ\ by randomising each data point within its uncertainty error
bar (errors are assumed normal with standard deviation equal to the
size of the error bar). Finally, we find the best (minimum $\chi^2$)
model for each bootstrapped version lightcurve and thus obtain the
distribution of best-fit parameters.

Figure \ref{fig:EllipsoidLightcurveFit} shows the best-fitting Jacobi
ellipsoid lightcurve and Figure \ref{fig:EllipsoidModel} shows the
corresponding model shape. The Monte Carlo distributions of best-fit
parameters are shown in Figure
\ref{fig:EllipsoidParameterDistribution}. Figures
\ref{fig:BinaryLightcurveFit}, \ref{fig:BinaryModel}, and
\ref{fig:BinaryParameterDistribution} show the best-fitting
lightcurve, best model shape and parameter distributions for Roche
binary models. Table \ref{table:FitParameters} summarises the best-fit
parameters for each model. The Roche binary fits are generally better
with a typical $\chi^2\approx 1.62$ per degree of freedom compared to
$\chi^2\approx 1.74$ per degree of freedom for the Jacobi ellipsoid
models. The Roche binary model successfully fits the different minima
in the lightcurve of \SQ, unlike the Jacobi ellipsoid model. The mean
scattering parameter for Jacobi ellipsoid fits is $k=0.1$, consistent
with a low-albedo surface, while for Roche binaries we find a mean
$k=0.4$, lending almost equal weights to (dark) lunar- and (bright)
icy-type terrains. Higher $k$ values imply stronger limb darkening,
which is needed to fit the different lightcurve minima.

\begin{figure}
  \includegraphics[width=\columnwidth]{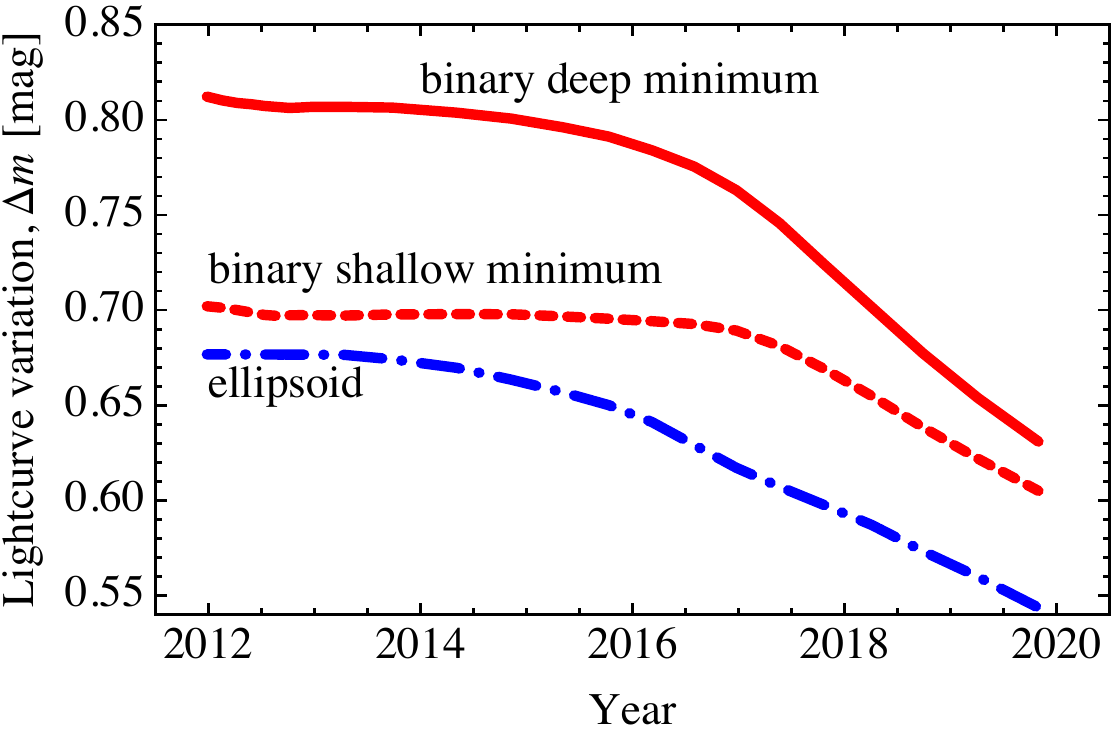}
  \centering

  \caption{Model lightcurve variation, $\Delta m$, as a function of
    time for the Jacobi ellipsoid (dash-dot, blue) and the Roche
    binary solution. In the case of the binary model, the changes for
    the shallow (dashed, red) and deep (solid, red) minima are shown.
    The change in $\Delta m$ plotted here is maximal as it assumes
  that the models have 90\degr\ obliquity. }

  \label{fig:VariationChange}
\end{figure}

\subsection{Bulk Density}

In \S\ref{sec:Shape} we found the Jacobi ellipsoid and Roche binary
that best fit the lightcurve of \SQ. Because these models assume
hydrostatic equilibrium, their shapes are uniquely related to bulk
density and spin period and allow us to use the latter to constrain
the former. Each model shape is a function of the dimensionless
parameter $\Omega^2=\omega^2/\left(\pi\,G\,\rho\right)$ where
$\omega=2\pi/P$ is the angular rotation frequency ($P$ is the period),
$G$ is the gravitational constant and $\rho$ is the bulk density. For
a spin period $P=7.21$ hr, the density is then calculated as
$\rho=280/\Omega^2$ kg m$^{-3}$.

Predictably, the two types of model imply very different bulk
densities (Figures \ref{fig:EllipsoidParameterDistribution} and
\ref{fig:BinaryParameterDistribution}, Table
\ref{table:FitParameters}). The Jacobi model fit yields
$\Omega^2=0.325\pm0.012$ and indicates a bulk density $\rho=860\pm30$
kg m$^{-3}$, consistent with an icy composition. The Roche binary
model has $\Omega^2=0.105\pm0.004$ and leads to $\rho=2700\pm100$ kg
m$^{-3}$ suggesting a rock-rich bulk composition for \SQ.

\section{Discussion} \label{sec:Discussion}


As the analysis in \S\ref{sec:Lightcurve} shows, \SQ\ oscillates in
brightness by $\Delta m=0.85\pm 0.05$ mag, making it the second most
variable KBO known, only surpassed by 2001$\,$QG$_{298}$ (hereafter
\QG) with $\Delta m=1.14\pm0.04$ mag \citep{2004AJ....127.3023S}.  For
bodies in hydrostatic equilibrium, lightcurve variation $\Delta m>0.9$
mag can only plausibly be explained by a tidally distorted, binary
shape \citep{1980Icar...44..807Weidenschilling, 1984A&A...140..265L}.
That is the case of \QG\ which was sucessfully modelled as a Roche
binary leading to an estimated bulk density near 660 kg m$^{-3}$
\citep{2004EP&S...56..997Takahashi, 2007AJ....133.1393Lacerda,
2010ApJ...719.1602Gnat}. The Roche binary model received further
support as \QG\ was re-observed in 2010 to show a predicted decrease
in variability to $\Delta m=0.7$ magnitudes, which allowed the
obliquity of the system to be estimated at very near 90$\degr$
\citep{2011AJ....142...90Lacerda}.

With $\Delta m=0.85$ mag and $P=7.21$ hr, \SQ\ lies at the threshold
between the Jacobi and Roche sequences \citep{1984A&A...140..265L,
2004AJ....127.3023S}. Indeed, we find that \SQ\ can be fitted
reasonably well both by Jacobi and Roche models. However, one
important feature of the lightcurve, the asymmetric lightcurve minima,
is only naturally fitted by the Roche binary model. The Jacobi model
misses the data points that mark the faintest point of the lightcurve
(Figure \ref{fig:EllipsoidLightcurveFit}). In theory, a special
arrangement of brighter and darker surface patches could be adopted to
ensure that the Jacobi model fit the fainter lightcurve minimum.
However, the binary model does not require any further assumptions and
thus provides a simpler explanation for the asymmetric lightcurve
minima. Figure \ref{fig:VariationChange} plots the change in
lightcurve variation, $\Delta m$, for both models. To maximise the
change, we assumed the models to have 90\degr\ obliquity so that an
angular displacement, $\nu$, along the heliocentric orbit will
translate into a change in aspect angle $\theta\approx\nu$
\citep{2011AJ....142...90Lacerda}. The two models produce slightly
different behaviour and future observations may help rule out one of
the solutions.


The Jacobi and Roche model solutions predict significantly different
bulk densities for \SQ. The former is consistent with a density around
860 kg m$^{-3}$ which would indicate a predominantly icy interior and
significant porosity. The Roche binary model implies a density close
to 2670 kg m$^{-3}$, consistent with a rocky bulk composition.
Densities higher than 2000 kg m$^{-3}$ have only been measured for the
larger KBOs, Eris \citep{2007Sci...316.1585Brown,
2011Natur.478..493Sicardy}, Pluto \citep{1993AJ....105.2319Null,
2006AJ....132..290Buie}, Haumea \citep{2006ApJ...639.1238Rabinowitz,
2007AJ....133.1393Lacerda} and Quaoar
\citep{2013A&A...555A..15Fornasier}. Densities for objects with
diameters similar to \SQ\ tend to fall in the range $500<\rho<1000$ kg
m$^{-3}$ \citep{2012Icar..220...74Grundy,
2012Icar..219..676Stansberry} with the possible exception of (88611)
Teharonhiawako with $\rho\approx1400$ kg m$^{-3}$
\citep{2003EM&P...92..409Osip, 2013A&A.Lellouch}. If confirmed, the
Roche model density of \SQ\ would make it one of the highest density
known KBOs and the densest of its size.

In a scenario in which the Haumea family was produced by a collision
that ejected the volatile-rich mantle of the proto-Haumea, its members
would be expected to be mainly icy in composition, with high albedo
surfaces. The Jacobi model density for \SQ\ would favour such a
scenario (although the implied surface scattering is inconsistent with
an icy, high albedo and high limb-darknening surface) whereas the
Roche model density would be harder to explain in the context of the
family. Haumea's density $\rho\approx2600$ kg m$^{-3}$ and water-ice
spectrum implies a rocky core surrounded by a veneer of ice. If \SQ\
has high-density and was produced from a collision onto Haumea then it
must be a fragment from the core material.


Broadband near-infrared photometry of \SQ\ suggests a surface rich in
water ice \citep{2010A&A...511A..72Snodgrass}. Our measurements
indicate a nearly solar\footnote{$(B-R)_\mathrm{\sun}=1.00\pm0.02$
mag \citep{2006MNRAS.367..449Holmberg}} surface colour
$B-R=1.05\pm0.18$ mag and a steep (although poorly constrained) phase
function with slope $\beta=0.95\pm0.41$ mag/$\degr$. While the visible
and infrared colours of \SQ\ match those of other members of the
Haumea family, the phase function is much steeper.

The albedo of \SQ\ is unknown. Although its surface is blue and
possibly water-ice rich, these properties do not necessary imply high
albedo. For instance, 2002$\,$MS$_{4}$ has blue colour
\citep[$B-R\approx1.0$ mag;][]{2012A&A...546A..86Peixinho} but low
albedo \citep[$p_V\approx0.05$;][]{2013A&A.Lellouch}, and Quaoar
displays strong water-ice absorption \citep{2004Natur.432..731Jewitt}
despite its relatively dark
\citep[$p_V\approx0.12$;][]{2013A&A.Lellouch} and red surface
\citep[$B-R\approx1.6$ mag;][]{{2012A&A...546A..86Peixinho}}. Water
ice has been spectroscopically detected on objects with albedos as low
as 0.04, e.g.\ Chariklo \citep{2009A&A...501..777Guilbert}, and as
high as 0.80, e.g.\ Haumea \citep{2007ApJ...655.1172T}. 

The shape models and density estimates presented above assume an
idealised fluid object in hydrostatic equilibrium. As a limiting case,
the simplification is useful because it offers a simple and unique
relation between shape, spin period and bulk density. However, \SQ\ is
a solid body and likely behaves differently.
\citet{2001Icar..154..432Holsapple, 2004Icar..172..272Holsapple} have
studied extensively the equilibrium configurations of rotating solid
bodies---sometimes termed ``rubble piles''---that possess no tensile
strength but that can retain shapes bracketing the hydrostatic
solution due to pressure-induced, internal friction. Similar studies
were performed for Roche figures of equilibrium by
\citet{2009Icar..200..636Sharma}. The deviation from the hydrostatic
equilibrium solution is usually quantified in terms of an increasing
angle of friction, $0\mathrm{\degr}<\phi<90\mathrm{\degr}$. For a
positive value of $\phi$, a range of bulk densities (which includes
the hydrostatic equilibrium solution) is possible for an object with a
given shape and spin rate. 

For a plausible range of albedos, \SQ\ has an equivalent diameter in
the range $150<D<450$ km. The giant planet icy moons in the same size
range (Amalthea at Jupiter and Mimas, Phoebe and Janus at Saturn;
Hyperion has chaotic rotation and is ignored) lie at the threshold
between near hydrostatic shapes and slightly more irregular
configurations \citep{2010Icar..208..395Thomas,
2012Icar..219...86Castillo-Rogez}.  When approximated by triaxial
ellipsoids and plotted on the diagrams of
\citet{2001Icar..154..432Holsapple}, the shapes, spins and densities
of these moons are consistent with angles of friction $\phi<5\degr$
\citep[see also][]{2009Icar..200..636Sharma}. Similar values of $\phi$
are found for most large, approximately triaxial asteroids
\citep{2009Icar..200..304Sharma}. If we take our Jacobi ellipsoid
solution for \SQ\ and assume an angle of friction $\phi=5\degr$ then
we find that its density should lie in the range $670<\rho<1100$
kg~m$^{-3}$, i.e.\ a 30\% departure from the idealised hydrostatic
equilibrium solution. A similar uncertainty applied to the Roche
binary density estimate yields a range $2050<\rho<3470$ kg~m$^{-3}$.

\section{Summary}

We present time-resolved photometric observations of Kuiper belt
object \SQlong\ obtained in August and October 2011 to investigate its
nature. Our results can be summarised as follows:

\begin{enumerate}

  \item \SQ\ exhibited a highly variable photometric lightcurve with a
    peak-to-peak range $\Delta m=0.85\pm0.05$ magnitudes and period
    $P=7.210\pm0.001$ hours. The object has an almost solar broadband
    colour $B-R=1.05\pm0.18$ mag, making it one of the bluest KBOs
    known. The phase function of \SQ\ is well matched by a linear
    relation with intercept $m_R(1,1,0)=5.52\pm0.36$ mag and slope
    $\beta=0.96\pm0.41$ mag/$\degr$. This linear phase function is
    consistent with an earlier measurement obtained in 2008 at phase
    angle $\alpha=0.62\degr$.

  \item The lightcurve implies that \SQ\ is highly elongated in shape.
    Assuming that the object is in hydrostatic equilibrium, we find
    that the lightcurve of \SQ\ is best fit by a compact Roche binary
    model with mass ratio $q\sim0.3$, and triaxial primary and
    secondary components with axes ratios $B/A\sim0.9$, $C/A\sim0.8$
    and $b/a\sim c/a\sim0.5$, separated by $d\sim1.1(A+a)$.  The data
    are also adequately fitted by a highly elongated, Jacobi triaxial
    ellipsoid model with axes ratios $B/A\sim0.55$ and $C/A\sim0.41$.
    Observations in this decade may be able to rule out one of the two
    solutions.

  \item If \SQ\ is a Roche binary then its bulk density is
    approximately 2670 kg $^{-3}$. This model-dependent density
    implies rock-rich composition for this object. However, if \SQ\ is
    a Jacobi ellipsoid we find a significantly lower density,
    $\rho\approx860$ kg m$^{-3}$ consistent with an icy, porous
    interior. These density estimates become uncertain at the 30\%
    level if we relax the hydrostatic assumption and account for
    ``rubble pile''-type configurations.

\end{enumerate}

\acknowledgments

We thank David Jewitt for comments, and Colin Snodgrass for supplying
an IRAF fringe removal routine optimised for EFOSC2. The presented data
were obtained at the ESO facilities at La Silla under programmes
087.C-0980A and 088.C-0634A.

\end{document}